\documentstyle[12pt]{article}
\begin{document}

\begin{flushright}
EDO-EP-50\\
DFPD 05/TH/17\\
June, 2005\\
hep-th/0506054
\end{flushright}
\vspace{20pt}

\pagestyle{empty}
\baselineskip15pt

\begin{center}
{\large\bf Worldline Approach of Topological BF Theory
\vskip 1mm
}

\vspace{20mm}

Ichiro Oda
          \footnote{
          E-mail address:\ ioda@edogawa-u.ac.jp
                  }
\\

\vspace{5mm}
          Edogawa University,
          474 Komaki, Nagareyama City, Chiba 270-0198, Japan\\

\vspace{5mm}

and

\vspace{5mm}

Mario Tonin
          \footnote{
          E-mail address:\ mario.tonin@pd.infn.it
                  }
\\
\vspace{5mm}
          Dipartimento di Fisica, Universita degli Studi di Padova,\\
          Instituto Nazionale di Fisica Nucleare, Sezione di Padova,\\
          Via F. Marzolo 8, 35131 Padova, Italy\\

\end{center}


\vspace{5mm}
\begin{abstract}
We present a worldline description of topological non-abelian BF theory 
in arbitrary space-time dimensions. It is shown that starting with a trivial 
classical action defined on the worldline, the BRST cohomology has
a natural representation as the sum of the de Rham cohomology. 
Based on this observation, we construct a second-quantized action of
the BF theory. Interestingly enough, this theory naturally gives us a minimal 
solution to the Batalin-Vilkovisky master equation of the BF theory. 
Our formalism sheds some light not only on an interplay between 
the Witten-type and the Schwarz-type topological quantum field theories 
but also on the role of the Batalin-Vilkovisky antifields and ghosts as 
geometrical and elementary objects.  
\end{abstract}
\newpage
\pagestyle{plain}
\pagenumbering{arabic}

\section{Introduction}

In recent years, we have seen an interesting progress on a covariant
quantization of Green-Schwarz superstring theories, which was a long-standing
problem for about twenty years since appearance of a paper by
Green and Schwarz \cite{GS}, by using pure spinors \cite{Ber1}. (See
also related papers \cite{Many}.) The formalism depends on a very
simple form of the BRST charge $Q_B = \oint \lambda^\alpha d_\alpha$,
where $\lambda^\alpha$ is a pure spinor satisfying the pure spinor
equation $\lambda^\alpha \Gamma^a_{\alpha\beta} \lambda^\beta = 0$
and $d_\alpha$ is a spinorial covariant derivative.
 
In a lecture note introducing the pure spinor formalism of superparticle
and superstring \cite{Ber2}, Berkovits has discussed that the 10D super Yang-Mills
theory can be obtained through the BRST quantization of a superparticle
action involving pure spinors, just as the 3D Chern-Simons theory which
is in essence a topological theory, can be obtained via the BRST quantization
of a particle action \cite{Warren}. 
This worldline description of the 3D Chern-Simons theory was 
gained by dimensionally reducing a worldsheet action for a Chern-Simons string theory
by Witten \cite{Witten1}\footnote{A supersymmetric extension was considered in 
\cite{Grassi}.}  to a worldline action and was introduced as 
just a prelude to the pure spinor formalism, but the worldline description is 
of interest in its own right from some reasons mentioned below.

For instance, it is nowadays well known that there are two types of topological 
quantum field theories. One is called the Witten-type topological quantum
field theories, or the cohomological type where the classical action is
some topological invariants or simply zero \cite{Witten2, Baulieu}. 
This type of topological quantum field theories was originally introduced 
to understand Donaldson invariants defined on 4D differentiable manifolds. 
The other is sometimes called the Schwarz-type topological quantum field theories, 
for which the classical action has the form of Chern-Simons action or 
BF action \cite{Blau, Horowitz, Wallet}. This type of topological quantum field theories was 
originally developed not only to make possible the quantization of linear $p$-form theory 
but also to formulate the Ray-Singer analytical torsions of the de Rham complex 
in a field theoretic language \cite{Schwarz}. 
These two types of topological quantum field theories share the property 
that their partition functions are independent of the metric and 
that the only observables are topological invariants
of the underlying space-time manifold, but appear to be disconnected as a field
theory since the Schwarz-type has a nontrivial classical action while the Witten-type
has a trivial action at least classically and therefore possesses a well-known topological
symmetry. The worldline approach which we wish to investigate in this paper bridges 
the gap between the two types of topological field theories to some extent, which is 
one reason behind the motivations of the present paper.

Another interesting reason of the worldline approach is that we can construct 
a second-quantized theory of the Schwarz-type topological quantum field theories 
by using the BRST charge of the worldline action in a natural way. This second-quantized
action includes the Batalin-Vilkovisky antifields in additon to a tower of ghosts
and is found to be a minimal solution to the BV master equation, which is a heart of 
the Batalin-Vilkovisky quantization algorithm \cite{Batalin}. 
Thus, the formalism at hand gives us a different standpoint of the Batalin-Vilkovisky 
algorithm and antifields, and also provides us a geometrical origin of the antifields.

In section 2, from the viewpoint of the canonical quantization, we analyze the BRST cohomology 
of the worldline approach in detail in order to explain why its BRST cohomology can describe 
the moduli space of the Schwarz-type topological quantum field theories. In section 3, we
construct a second-quantized theory corresponding to BF theory. Section 4
is devoted to discussions.

\section{Review of the Worldline Approach}

To begin with, we shall investigate the worldline approach for topological field theories 
developed by Berkovits \cite{Ber2} in some detail from a slightly different perspective. 
The theory involves $D = n + 2$ bosonic variables $x^\mu$ representing the position 
of a particle and their canonical conjugate momenta $P_\mu$ in addition to the Lagrange
multipliers $l^\mu$ imposing the constraints $P_\mu \approx 0$. In this article,
we limit ourselves to consider the cases $D \geq 2$, i.e., $n \geq 0$. The classical
action of the system is of form 
\begin{eqnarray}
S_c = \int d \tau L_c = \int d \tau (\dot{x}^\mu P_\mu + l^\mu P_\mu),
\label{2.1}
\end{eqnarray}
where $\dot{x}^\mu = \frac{\partial x^\mu}{\partial \tau}$. 

Here let us quantize the system in a canonical manner. The canonical conjugate
momenta for $x^\mu$ and $l^\mu$ are given by
\begin{eqnarray}
P_\mu &=& \frac{\delta S_c}{\delta \dot{x}^\mu}, \nonumber\\
\pi_\mu &=& \frac{\delta S_c}{\delta \dot{l}^\mu} \approx 0.
\label{2.2}
\end{eqnarray}
The latter equation gives rise to the primary constraints. The Hamiltonian
$H$ is then defined as
\begin{eqnarray}
H = P_\mu \dot{x}^\mu + \pi_\mu \dot{l}^\mu - L_c = - l^\mu P_\mu.
\label{2.3}
\end{eqnarray}
{}From this Hamiltonian and the primary constraints, one obtains the
secondary constraints
\begin{eqnarray}
P_\mu \approx 0.
\label{2.4}
\end{eqnarray}
Using the secondary constraints, the Hamiltonian $H$ is weakly zero,
so we have no more constraints. At the same time, the secondary
constraints make the classical action vanish, thereby implying a 
topological nature of the classical system\footnote{This can
be also seen more directly by performing the path integral over $l^\mu$ 
in the action (\ref{2.1}) and then over $P_\mu$.}, that is, this system is
an instance of the Witten-type topological field theories. 
It turns out that the primary and secondary
constraints constitute the first-class constraints whose generator 
is given by
\begin{eqnarray}
G = \int d \tau ( - \dot{\varepsilon}_\mu \pi^\mu 
+ \varepsilon_\mu P^\mu ).
\label{2.5}
\end{eqnarray}
Actually the generator yields the topological symmetry 
\begin{eqnarray}
\delta x^\mu &=& - \varepsilon^\mu, \nonumber\\
\delta l^\mu &=& \dot{\varepsilon}^\mu, \nonumber\\
\delta P_\mu &=& 0,
\label{2.6}
\end{eqnarray}
under which the classical action (\ref{2.1}) is manifestly invariant.

Now let us move on to the BRST quantization. The topological symmetry
gives rise to the following BRST transformation:
\begin{eqnarray}
\delta_B x^\mu &=& - c^\mu, \nonumber\\
\delta_B l^\mu &=& \dot{c}^\mu, \nonumber\\
\delta_B P_\mu &=& 0, \nonumber\\
\delta_B b_\mu &=& B_\mu, \nonumber\\
\delta_B c^\mu &=& \delta_B B_\mu = 0.
\label{2.7}
\end{eqnarray}
Since we adopt the gauge conditions for the topological symmetry
\begin{eqnarray}
l^\mu + \frac{1}{2} P^\mu = 0,
\label{2.8}
\end{eqnarray}
the gauge-fixed, BRST-invariant action is given by
\begin{eqnarray}
S &=& \int d \tau \Big[\dot{x}^\mu P_\mu + l^\mu P_\mu 
+ \delta_B \Big( b_\mu ( l^\mu + \frac{1}{2} P^\mu ) \Big) \Big], \nonumber\\
&=& \int d \tau \Big[\dot{x}^\mu P_\mu + l^\mu P_\mu 
+ B_\mu ( l^\mu + \frac{1}{2} P^\mu ) + \dot{c}^\mu b_\mu \Big].
\label{2.9}
\end{eqnarray}
In order to simplify this quantum action further, we shall carry
out the path integral over the auxiliary fields $l^\mu$ and $B_\mu$
whose result is given by
\begin{eqnarray}
S = \int d \tau \Big[\dot{x}^\mu P_\mu - \frac{1}{2} P_\mu P^\mu 
+ \dot{c}^\mu b_\mu \Big].
\label{2.10}
\end{eqnarray}
The Noether theorem makes it possible to construct the BRST charge
\begin{eqnarray}
Q_B = i c^\mu P_\mu.
\label{2.11}
\end{eqnarray}
The action (\ref{2.10}) is also invariant under the scale transformation
\begin{eqnarray}
c^\mu &\rightarrow& e^\rho c^\mu, \nonumber\\
b_\mu &\rightarrow& e^{-\rho} b_\mu,
\label{2.12}
\end{eqnarray}
where $\rho$ is a real parameter, so we can define the ghost number
charge through the Noether theorem by
\begin{eqnarray}
Q_c = - i c^\mu b_\mu.
\label{2.13}
\end{eqnarray}
Since we have the commutators
\begin{eqnarray}
&{}&[ Q_c, c^\mu ] = c^\mu, \nonumber\\
&{}&[ Q_c, b_\mu ] = - b_\mu,
\label{2.14}
\end{eqnarray}
$c^\mu$ and $b_\mu$ have respectively $+1$ and $-1$ ghost numbers.
In the above, the (anti-)commutation relations are set up 
as usual by
\begin{eqnarray}
[ x^\mu, P_\nu ] &=& i \delta^\mu_\nu, \nonumber\\
\{ c^\mu, b_\nu \} &=& i \delta^\mu_\nu,
\label{2.15}
\end{eqnarray}
with the other (anti-)commutators vanishing.

The quantization is incomplete unless one fixes the concrete representation
of the above algebra. At this stage, we find that there is a natural 
representation which is nothing but the representation on the space of 
differential forms.
\begin{eqnarray}
{\cal{H}} = \Omega(M) = \bigoplus_{p=0}^D \Omega^p(M),
\label{2.16}
\end{eqnarray}
where $M$ and $\Omega^p(M)$ denote the $D$-dimensional manifold and 
the space of the $p$-forms, respectively.
On this space, the variables in the worldline approach are represented
as operators and have the following correspondence with differential
forms:
\begin{eqnarray}
x^\mu &\longleftrightarrow& x^\mu \otimes, \nonumber\\
P_\mu &\longleftrightarrow& -i \frac{\partial}{\partial x^\mu}, \nonumber\\
c^\mu &\longleftrightarrow& d x^\mu \wedge, \nonumber\\
b_\mu &\longleftrightarrow& i \times i_{\frac{\partial}{\partial d x^\mu}},
\label{2.17}
\end{eqnarray}
where $i_V$ means the interior product which is an operation producing
$(k-1)$-form from $k$-form by contracting the differential form with the
vector field $V$.
Then, the physical Hilbert space, which is denoted as $|\psi>$, should
be annihilated by the BRST charge
\begin{eqnarray}
Q_B |\psi> = 0.
\label{2.18}
\end{eqnarray}
In this representation, we also have the following correspondence:
\begin{eqnarray}
|0> &\longleftrightarrow& 1, \nonumber\\
c^\mu |0> &\longleftrightarrow& d x^\mu, \nonumber\\
\cdots &\longleftrightarrow& \cdots, \nonumber\\
c^{\mu_1} \cdots c^{\mu_D} |0> &\longleftrightarrow& d x^{\mu_1} \wedge 
\cdots \wedge d x^{\mu_D}, \nonumber\\
Q_B &\longleftrightarrow& d x^\mu \wedge \frac{\partial}{\partial x^\mu}
= d, \nonumber\\
- H &\longleftrightarrow& \frac{1}{2} \bigtriangledown,
\label{2.19}
\end{eqnarray}
where $\bigtriangledown$ is the Laplacian operator. 
Here we have denoted by $|0>$ the vector annihilated by $Q_c$
in addition to $Q_B$, so $|0>$ belongs to the ghost number $0$
sector.\footnote{As an alternative interpretation of this representation,
one could regard $b_\mu$ as annihilation operators and $c^\mu$
as creation operators. Then, the whole physical state is constructed
by operating a finite number of creation operators $c^\mu$
on the "vacuum" $|0>$ which is destroyed by annihilation
operators $b_\mu$.}
Since $[ Q_c, c^\mu ] = c^\mu$ from (\ref{2.14}), the 
ghost number of a general state $c^{\mu_1} \cdots c^{\mu_p} |0>$
is identified with the form-degree $p$.
Also notice that the physical state condition leads to the zero energy condition, 
so the whole physical state is consisted of only the ground states, 
which are simply the harmonic forms. In other words, the BRST cohomology 
${\cal{H}}$ at hand is the direct sum of the de Rham cohomology group $H^p(M)$ 
of forms on the $D$-dimensional manifold $M$:
\begin{eqnarray}
{\cal{H}} = \bigoplus_{p=0}^D H^p(M).
\label{2.20}
\end{eqnarray}
Here recall that the space ${\cal{N}}$ of classical solutions of
the 3$D$ Chern-Simons theory (and the BF theory in arbitrary dimensions) is
a finite dimensional de Rham cohomology group (and the direct sum
of two de Rham cohomology groups) \cite{Blau}. Hence, it is natural to expect that 
the moduli space of the Schwarz-type of topological field theories might
be described in terms of the BRST cohomology of the worldline approach 
mentioned above, which is of the Witten-type \cite{Witten2, Baulieu}.
In fact, Berkovits has shown that this is indeed the case for the
3$D$ Chern-Simons theory \cite{Ber2}. One of motivations behind the present article
is show explicitly that this holds for the BF theory \cite{Blau, Horowitz, Wallet}
in arbitrary dimensions as well. 
It is worth noting that the BF theory in more than three space-time 
dimensions has an on-shell reducible symmetry in addition to the usual Yang-Mills
gauge symmetry, so our application of the worldline approach to the BF
theory is not so obvious as we consider naively. 

To close this section, it is valuable to point out that the $b$ ghost
associated to the $\tau$-reparametrization symmetry can be found as follows:
The Hamiltonian $H$ is rewritten as
\begin{eqnarray}
H = - l^\mu P_\mu = \frac{1}{2} P_\mu^2.
\label{2.21}
\end{eqnarray}
The fundamental equation which the $b$ ghost must satisfy is \cite{Witten1}
\begin{eqnarray}
\{ Q_B, b \} = H.
\label{2.22}
\end{eqnarray}
Hence, we can make the $b$ ghost by
\begin{eqnarray}
b = - \frac{1}{2} b_\mu P^\mu.
\label{2.23}
\end{eqnarray}
This form of the $b$ ghost is physically reasonable since the reparametrization
is part of more huge topological symmetry. Note that this $b$ ghost
is a composite field as that in the pure spinor formalism.

\section{Worldline Description of BF theory}

In this section, on the basis of the observation in previous section, 
we wish to present a worldline description of BF theory in a general
space-time dimension. Since the path of argument is similar in both
the abelian and the non-abelian gauge groups, we shall discuss only
the case of the non-abelian gauge group, from which we can extract 
the abelian BF theory in a straightforward way.

The classical action of the BF theory in $D = n + 2 \geq 2$ space-time
dimensions takes the form\footnote{The wedge product among forms is
always understood.}
\begin{eqnarray}
S = \int_{M_D} Tr ( B F ),
\label{3.1}
\end{eqnarray}
where $A$ indicates a Lie algebra valued $1$-form connection and $F$
is its curvature $2$-form defined by $F = dA + A^2$, and $B$ is a
section belonging to $\Omega^n (M)$, i.e., a Lie algebra valued $n$-form.
The equations of motion from this action read
\begin{eqnarray}
F &=& 0, \nonumber\\
DB &=& 0,
\label{3.2}
\end{eqnarray}
where the covariant derivative is defined as $D = d + [A, \ ]$. 
The action is invariant under the conventional Yang-Mills gauge transformation 
with the $0$-form gauge parameter $\varepsilon(x)$ and the non-abelian symmetry
associated with $B$ field with the $(n-1)$-form gauge parameter $\lambda(x)$
\begin{eqnarray}
\delta A &=& D \varepsilon, \nonumber\\
\delta B &=& [ B, \varepsilon ] + D \lambda.
\label{3.3}
\end{eqnarray}
When one attempts to quantize this system, a well-known complication
appears owing to the latter transformation in four and higher space-time
dimensions. Namely, the non-abelian symmetry for $B$ field is on-shell
reducible in the sense that $\lambda = D \lambda'$ with $\lambda'$ being
any $(n-2)$-form becomes the zero modes from the equation of motion
$F = 0$. (Note that such a sequence of reducible symmetries exists
until $\lambda'$ is a $0$-form.) In order to quantize such the on-shell
reducible theory, one might rely on the Batalin-Vilkovisky algorithm
\cite{Wallet}.

Now we are ready to present a worldline description of the above non-abelian
BF theory in $D = n + 2$ space-time dimensions. Before doing that, let us
first recall the result in the previous section that the BRST cohomology
of the worldline approach is ${\cal{H}} = \bigoplus_{p=0}^{n+2} H^p(M)$
whereas the moduli space of the BF theory is 
${\cal{N}} = H^n(M) \oplus H^1(M)$ \cite{Blau}.\footnote{Precisely speaking, 
this is true only when the gauge group is the abelian group. In the present case,
we have a non-abelian generalization of it since there is the non-linear
nilpotent operator $D$ from the equations of motion $F = 0$.}
Thus, if we want to make ${\cal{N}}$ coincide with ${\cal{H}}$, it is necessary
to add the missing de Rham cohomology groups to ${\cal{N}}$. Actually,
such the cohomology groups are precisely supplied by a tower of reducible
ghosts and the Yang-Mills ghost as well as the Batalin-Vilkovisky antifields
as we will show shortly.

Next, corresponding to $A$ and $B$ fields in the BF theory, let us introduce
two kinds of functionals $\Psi(c, x)$ with ghost number $1$ and $\Phi(c, x)$
with ghost number $n$, and then expand them in the powers of $c^\mu$ which are
the topological ghosts in the worldline action of a particle
\begin{eqnarray}
\Psi(c, x) &=& C(x) + c^\mu A_\mu(x) + \frac{1}{2} c^{\mu_1} c^{\mu_2}
B^*_{\mu_1\mu_2}(x) + \cdots + \frac{1}{(n+2)!} c^{\mu_1} \cdots c^{\mu_{n+2}}
B^*_{\mu_1 \cdots \mu_{n+2}}(x), \nonumber\\
\Phi(c, x) &=& B(x) + c^\mu B_\mu(x) + \frac{1}{2} c^{\mu_1} c^{\mu_2}
B_{\mu_1\mu_2}(x) + \cdots + \frac{1}{n!} c^{\mu_1} \cdots c^{\mu_n}
B_{\mu_1 \cdots \mu_n}(x) \nonumber\\
&{}& + \frac{1}{(n+1)!} c^{\mu_1} \cdots c^{\mu_{n+1}} A^*_{\mu_1 \cdots \mu_{n+1}}(x)
+ \frac{1}{(n+2)!} c^{\mu_1} \cdots c^{\mu_{n+2}} C^*_{\mu_1 \cdots \mu_{n+2}},
\label{3.4}
\end{eqnarray}
where the expanded terms terminate at a finite stage since $c^\mu$ are 
anticommuting and the space-time dimension is now $D = n + 2$.
Then, we can propose the following second-quantized action which is a 
natural generalization of the BF theory:
\begin{eqnarray}
S = \int_{M_D} d^D x d^D c \ Tr \Phi ( Q_B \Psi + \Psi \Psi ),
\label{3.5}
\end{eqnarray}
where $Q_B$ is the BRST charge of the worldline theory. Here the ghost
measure is defined as $\int d^D c \ c^1 \cdots c^D = 1$. This action is
invariant under the gauge transformations 
\begin{eqnarray}
\delta \Psi &=& Q_B \Omega + [ \Psi, \Omega ], \nonumber\\
\delta \Phi &=& [ \Phi, \Omega ] + Q_B \Lambda + [ \Psi, \Lambda ],
\label{3.6}
\end{eqnarray}
with $\Omega$ and $\Lambda$ having $0$ and $n-1$ ghost numbers, respectively.
Henceforth, we shall use the following notation: we define the total degree 
of arbitrary forms by the sum of the form degree and ghost number. The grading 
is then determined by the total degree and the square bracket means the commutator 
or anti-commutator depending on the grading. Concretely, we have the definition
$[ X, Y ] = XY - (-)^{xy} YX$ for a form $X$ with the total degree $x$ and
a form $Y$ with the total degree $y$. The topological ghosts $c^\mu$ are assumed
to have the total degree $1$. Moreover, we define a general $p$-form by
$A = \frac{1}{p!} dx^{\mu_1} \cdots dx^{\mu_p} A_{\mu_1 \cdots \mu_p}$,
instead of the conventional definition 
$A = \frac{1}{p!} A_{\mu_1 \cdots \mu_p} dx^{\mu_1} \cdots dx^{\mu_p}$,
since our definition is consistent with the expansion (\ref{3.4}). 
Otherwise, we would have numerous ugly factors of signature in the action
and the gauge transformations in components.

The action (\ref{3.5}) leads to the equations of motion
\begin{eqnarray}
Q_B \Psi + \Psi \Psi &=& 0, \nonumber\\
Q_B \Phi + [ \Psi, \Phi ] &=& 0.
\label{3.7}
\end{eqnarray}
Substituting the functionals (\ref{3.4}) into the action (\ref{3.5}),
we have an action for each component field by integrating over the
topological ghosts
\begin{eqnarray}
S &=& \int_{M_D}  \ Tr \Big[ B_{(n,0)} \Big( F_{(2,0)} 
+ [ C_{(0,1)}, B^*_{(2,-1)} ] \Big) + A^*_{(n+1,-1)} DC_{(0,1)}
\nonumber\\ 
&+& C^*_{(n+2,-2)}C_{(0,1)}C_{(0,1)} + \sum_{p=1}^n B_{(n-p,p)} 
\Big( DB^*_{(p+1,-p)} + [ C_{(0,1)}, B^*_{(p+2,-p-1)} ] \nonumber\\
&+& \frac{1}{2} \sum_{p'=0}^{p-2} [ B^*_{(p'+2,-p'-1)}, B^*_{(p-p',-p+p'+1)} ] 
\Big) \Big],
\label{3.8}
\end{eqnarray}
where we have put subscripts on fields in order to indicate the form-degree
and ghost number such that $X_{(p,q)}$ in general implies the field $X$
with $p$-form and ghost number $q$.

Moreover, provided that we also expand the gauge parameter functionals
$\Omega$ and $\Lambda$ in terms of $c^\mu$ as
\begin{eqnarray}
\Omega(c, x) &=& \omega(x) + c^\mu \omega_\mu(x) + \frac{1}{2} c^{\mu_1} c^{\mu_2}
\omega_{\mu_1\mu_2}(x) + \cdots + \frac{1}{(n+2)!} c^{\mu_1} \cdots c^{\mu_{n+2}}
\omega_{\mu_1 \cdots \mu_{n+2}}(x), \nonumber\\
\Lambda(c, x) &=& \lambda(x) + c^\mu \lambda_\mu(x) + \frac{1}{2} c^{\mu_1} c^{\mu_2}
\lambda_{\mu_1\mu_2}(x) + \cdots 
+ \frac{1}{(n+2)!} c^{\mu_1} \cdots c^{\mu_{n+2}} \lambda_{\mu_1 \cdots \mu_{n+2}},
\label{3.9}
\end{eqnarray}
then the gauge transformations are explicitly written out in components
(for $2 \leq p \leq n + 2$ and $1 \leq q \leq n$)
\begin{eqnarray}
&{}& \delta C_{(0,1)} = [ C_{(0,1)}, \omega_{(0,0)} ], \nonumber\\
&{}& \delta A_{(1,0)} = D \omega_{(0,0)} + [ C_{(0,1)}, \omega_{(1,-1)} ], \nonumber\\
&{}& \delta B^*_{(p,-p+1)} = D \omega_{(p-1,-p+1)} + [ C_{(0,1)}, \omega_{(p,-p)} ]
+ \sum_{p'=2}^{p} [ B^*_{(p-p'+2,-p+p'-1)}, \omega_{(p'-2,-p'+2)} ], \nonumber\\
&{}& \delta B_{(0,n)} = [ B_{(0,n)}, \omega_{(0,0)} ] + [ C_{(0,1)}, \lambda_{(0,n-1)} ], 
\nonumber\\
&{}& \delta B_{(q,n-q)} = D \lambda_{(q-1,n-q)} + [ C_{(0,1)}, \lambda_{(q,n-q-1)} ] 
+ \sum_{q'=0}^{q} [ B_{(q-q',n-q+q')}, \omega_{(q',-q')} ] \nonumber\\ 
&{}& + \sum_{q'=0}^{q-2} [ B^*_{(q-q',-q+q'+1)}, \lambda_{(q',n-q'-1)} ], \nonumber\\
&{}& \delta A^*_{(n+1,-1)} = D \lambda_{(n,-1)} + [ C_{(0,1)}, \lambda_{(n+1,-2)} ]  
+ [ A^*_{(n+1,-1)}, \omega_{(0,0)} ] \nonumber\\
&+& \sum_{p'=0}^{n} [ B_{(n-p',p')}, \omega_{(p'+1,-p'-1)} ] 
+ \sum_{p'=0}^{n-1} [ B^*_{(n-p'+1,-n+p')}, \lambda_{(p',n-p'-1)} ], 
\nonumber\\
&{}& \delta C^*_{(n+2,-2)} = D \lambda_{(n+1,-2)} + [ C_{(0,1)}, \lambda_{(n+2,-3)} ] 
+ [ A^*_{(n+1,-1)}, \omega_{(1,-1)} ] + [ C^*_{(n+2,-2)}, \omega_{(0,0)} ]
\nonumber\\
&+& \sum_{p'=0}^{n} [ B_{(n-p',p')}, \omega_{(p'+2,-p'-2)} ]
+ \sum_{p'=0}^{n} [ B^*_{(n-p'+2,-n+p'-1)}, \lambda_{(p',n-p'-1)} ].
\label{3.10}
\end{eqnarray}

A few comments are in order. One important comment is that the action (\ref{3.5}),
or equivalently (\ref{3.8}), turns out to be a minimal solution to the Batalin-Vilkovisky master
equation \cite{Batalin}. This is easily checked by a string field theoretic technique where
using the antibracket, the Batalin-Vilkovisky master equation is given by
\begin{eqnarray}
(S, S) &\equiv& \int d^D x d^D c \ Tr \frac{\delta S}{\delta \Psi} \frac{\delta S}{\delta \Phi}
\nonumber\\
&=& \int d^D x d^D c \ Tr (Q_B \Phi + [ \Psi, \Phi ])(Q_B \Psi + \Psi \Psi)
\nonumber\\
&=& \int d^D x d^D c \ Tr \Big( Q_B (\Psi Q_B \Phi + \Phi \Psi^2)
+ [ \Psi, \Phi ] \Psi^2 \Big) \nonumber\\
&=& 0,
\label{3.11}
\end{eqnarray}
where we have used the fact that $\int Q_B (\cdots) = 0$ and $\Psi$
is anticommuting.
We can rewrite this master equation to more familiar form for each component field
as follows:
\begin{eqnarray}
(S, S) &\equiv& \int d^D x \ Tr \Big( \frac{\partial^l S}{\partial C^*_{(0,-2)}} 
\frac{\partial^r S}{\partial C_{(0,1)}} + \frac{\partial^l S}{\partial A^*_{(1,-1)}} 
\frac{\partial^r S}{\partial A_{(1,0)}} 
+ \sum_{p=0}^{n} \frac{\partial^l S}{\partial B^*_{(p,p-n-1)}} 
\frac{\partial^r S}{\partial B_{(p,-p+n)}} \Big)
\nonumber\\
&=& 0,
\label{3.12}
\end{eqnarray}
where we have considered the dual fields for the Batalin-Vilkovisky antifields.

The second comment is that the second-quantized action (\ref{3.5}) is
automatically equipped with ghosts, ghosts of ghosts and antifields 
for reducible gauge symmetries for $B$ field in addition to the
Yang-Mills ghost and antifield. This in turn implies that we should take
account of the Batalin-Vilkovisky antifields on an equal footing with ghosts
in order to realize the huge second-quantized symmetry (\ref{3.6}) or (\ref{3.10}).
Of course, it is easy to recover the original BF action (\ref{3.1}) with the
gauge transformations (\ref{3.3}) from the second-quantized action (\ref{3.8})
through elimination of ghosts and the antifields by using the huge 
symmetry (\ref{3.10}) at least locally, but not globally. 
In the Batalin-Vilkovisky algorithm for quantization, 
recall that the action is a generator for the BRST transformation in the 
antibracket in the sense that $s X = (X, S)$ and the antifields must be 
gauge-fixed by selecting a suitable gauge fermion. 
Actually, using the equation $s X = (X, S)$, it is easy to 
derive the BRST transformation whose result is given by
(for $3 \leq p \leq n + 2$ and $1 \leq q \leq n$)
\begin{eqnarray}
&{}& s C_{(0,1)} = C_{(0,1)} C_{(0,1)}, \nonumber\\
&{}& s A_{(1,0)} = D C_{(0,1)}, \nonumber\\
&{}& s B^*_{(2,-1)} = - ( F_{(2,0)} + [ C_{(0,1)}, B^*_{(2,-1)} ] ), \nonumber\\
&{}& s B^*_{(p,-p+1)} = - ( D B^*_{(p-1,-p+2)} + [ C_{(0,1)}, B^*_{(p,-p+1)} ]
+ \frac{1}{2} \sum_{p'=0}^{p-4} [ B^*_{(p'+2,-p'-1)}, B^*_{(p-p'-2,-p+p'+3)} ] ), \nonumber\\
&{}& s B_{(0,n)} = [ C_{(0,1)}, B_{(0,n)} ], \nonumber\\
&{}& s B_{(q,n-q)} = D B_{(q-1,n-q+1)} + [ C_{(0,1)}, B_{(q,n-q)} ] 
+ \sum_{q'=0}^{q-2} [ B^*_{(q-q',-q+q'+1)}, B_{(q',n-q')} ], \nonumber\\
&{}& s A^*_{(n+1,-1)} = - ( D B_{(n,0)} + [ C_{(0,1)}, A^*_{(n+1,-1)} ] 
+ \sum_{p'=1}^{n} [ B^*_{(p'+1,-p')}, B_{(n-p',p')} ] ), \nonumber\\
&{}& s C^*_{(n+2,-2)} = - ( D A^*_{(n+1,-1)} + [ C_{(0,1)}, C^*_{(n+2,-2)} ] 
+ \sum_{p'=0}^{n} [ B^*_{(p'+2,-p'-1)}, B_{(n-p',p')} ] ).
\label{3.13}
\end{eqnarray}

The final remark is related to the extended differential calculus on the
universal bundle where the extended differential operator is the sum of the 
exterior derivative $d$ and the BRST transformation $s$ \cite{Baulieu}
\begin{eqnarray}
\tilde{d} = d + s,
\label{3.14}
\end{eqnarray}
and the universal $1$-form connection $\tilde{A}$ is the sum of the
gauge connection $A$ and the Yang-Mills ghost $C$
\begin{eqnarray}
\tilde{A} = A_{(1,0)} + C_{(0,1)},
\label{3.15}
\end{eqnarray}
since each object on the RHS carries the same total degree $1$.
It is true that this extended formalism yields the desired BRST 
transformation very nicely.\footnote{Here the "flatness" condition for
the extended curvature $2$-form \cite{Mario} plays an important role.} 
The use of the total degree also seems to
suggest that we could formulate the present theory in terms of the extended 
differential calculus on the universal bundle.
However, after some efforts we have found it difficult to formulate 
a second-quantized theory in a covariant manner in the framework of the 
extended differential calculus though we need more study to clarify 
this point in future.

\section{Discussions}

In this article, we have presented a worldline description of topological 
BF theory in arbitrary space-time dimensions and found that this formulation
provides a useful tool for obtaining a minimal solution to the Batalin-Vilkovisky 
master equation without solving it directly, which is usually a tough work especially
for the system with reducible on-shell symmetries.\footnote{See also related
works \cite{Dayi, Kawamoto}.}  In the second-quantized formalism, 
the ghosts and ghosts of ghost as well as the corresponding 
antifields are naturally required to participate in the action to realize
the gauge symmetry off-shell. In this sense, the antifields play the same
role as the ghosts and should be regarded as the geometrical and fundamental
objects in the construction of a second-quantized theory. It is
remarkable to notice that the missing de Rham cohomology groups 
in the original BF theory are neatly provided with such the fields
by taking account of general functionals.

It is natural to ask whether the present formulation can apply to the other
systems to get a minimal solution to the master equation. In this context, we
should pay attention to the form of the BRST charge $Q_B = i c^\mu P_\mu
= c^\mu \partial_\mu$ and the relation (\ref{2.22}). If we introduce
the dual BRST charge by $\tilde{Q}_B = -b_\mu P^\mu$, we can have a
suggestive equation $\frac{1}{2} \{ Q_B, \tilde{Q}_B \} = H$ and
as a result obtain the second-order Laplace operator. However, it seems to be
difficult to get a nilpotent operator associated to the Laplacian because
of the characteristic feature of the topological ghosts being space-time vectors, 
so the application of the present formulation might be limited to only
the system with the first-order differential operator in the
kinetic term in the action.

Finally, it is known that when we specify the space-time dimensions to three and
the gauge group to the $SO(1,2)$ group, the BF theory is reduced to 
three-dimensional gravity, which is essentially topological and 1-loop exact \cite{Oda}.
Thus, the BF theory at hand might be relevant to a topological gravity in three
dimensions or a closed string sector of superstring theory though we need much works 
to be done in future to render this idea realistic.

\begin{flushleft}
{\bf Acknowledgements}
\end{flushleft}

We are grateful to N. Berkovits for useful discussions and comments.
The work of the first author (I.O.) was partially supported by
the Grant-in-Aid for Scientific Research (C) No.14540277 from 
the Japan Ministry of Education, Science and Culture.
The work of the second author (M.T.) was  supported by the European
Community's Human Potential Programme under contract MRTN-CT-2004-005104 
"Constituents, Fundamental Forces and Symmetries of the Universe".

\newpage


\end{document}